\documentstyle[12pt]{article}
\newcommand{\pbarn}{\hbox {pb}}
\newcommand{\lumun}{{\hbox {pb}^{-1}}{\hbox {yr}^{-1}}}
\newcommand{\hc}{\hbox {h.c.}}
\newcommand{\re}{\hbox {Re}}
\newcommand{\im}{\hbox {Im}}
\newcommand{\mev}{\hbox {MeV}}
\newcommand{\gev}{\hbox {GeV}}
\newcommand{\etal}{\mbox{\em et al.}}
\newcommand{\prdj}[1]{{ \it Phys.~Rev.}~{\bf D{#1}}}
\newcommand{\prlj}[1]{{ \it Phys.~Rev.~Lett.}~{\bf {#1}}}
\newcommand{\plbj}[1]{{ \it Phys.~Lett.}~{\bf {#1B}}}

\newcommand{\ptpj}[1]{{ \it Prog.~Theor.~Phys.}~{\bf {#1}}}

\newcommand{\epem} {\mbox{$e^+e^-$}}
\newcommand{\ttbar}{\mbox{$t\bar{t}$}}
\newcommand{\apes} {\mbox{${\cal A}_{P}$}}
\newcommand{\tdec} {\mbox{$t\rightarrow W^+ b$}}
\newcommand{\tbardec} {\mbox{$\bar{t} \rightarrow W^- \bar{b}$}}
\newcommand{\wphel}{\mbox{$\lambda_{W^+}$}}
\newcommand{\wmhel}{\mbox{$\lambda_{W^-}$}}
\newcommand{\bhel} {\mbox{$h_{b}$}}
\newcommand{\thel} {\mbox{$h_{t}$}}
\newcommand{\ephel} {\mbox{$h_{e^+}$}}
\newcommand{\emhel} {\mbox{$h_{e^-}$}}
\newcommand{\ttprod} {\mbox{$\epem \rightarrow \ttbar$}}
\newcommand{\sstop}{\tilde t}

\begin{document}
\begin{flushright}
CERN-TH.6806/93
\end{flushright}
\vspace*{2cm}
\begin{center}
{\large{\bf CP Violation in {\boldmath $t \bar{t}$} Production
at {\boldmath $e^+ e^-$} Colliders} \\
\vspace*{2cm}
Bohdan Grz\c{a}dkowski\footnote{E-mail address: bohdang@cernvm.}\\
\vspace*{1cm}
       CERN, CH-1211 Geneva 23, Switzerland} \\
\vspace*{.5cm}
       and \\
\vspace*{.5cm}
{\large Institute for Theoretical Physics,\\
        University of Warsaw, \\
        Ho\.{z}a 69, PL-00-681 Warsaw, Poland.} \\
\vspace*{1cm}
{\bf Abstract} \\
\end{center}
The general production {\it and} decay mechanism of $\ttbar$
in future high-energy $\epem$ colliders has been investigated
in a model-independent way,
focusing on an observation of possible CP violation.
Angular asymmetries sensitive to CP violation either in the production
{\it or} in the subsequent decays
have been found. General
considerations are illustrated through 1-loop effects
induced
by gluino exchange in the supersymmetric version of the Standard Model.
\vspace*{4.5cm}
\begin{flushleft}
\parbox{2.in}
{CERN-TH.6806/93\\
February 1993}
\end{flushleft}
\setcounter{page}{0}
\thispagestyle{empty}
\newpage
\section{Introduction}

Linear high-energy $\epem$ colliders can prove to be
very useful laboratories
to study the physics of the top quark. One of the many interesting issues
one can investigate there is, for instance, whether or not
CP is violated in $\ttbar$ production and their subsequent decays.
Since at present very little is known experimentally about
the properties of the top quark, one can even expect there very
unconventional effects.
In spite of spectacular successes in experimental high-energy
physics (e.g.    precision tests of the Standard Model)
the origin of CP violation is still a mystery.
The standard theory of Kobayashi and Maskawa~\cite{km}
provides explicit CP violation through phases of Yukawa couplings;
however, many other attractive schemes of CP violation
may emerge in extensions of the Standard Model.
Those unconventional sources of CP violation can be
tested in $\ttbar$ production at future linear $\epem$ colliders.
For instance, CP violation induced by the neutral scalar
sector              in the two Higgs doublet standard
model~\cite{weinberg} could lead to observable effects~\cite{keung}
in $\ttprod$, provided sufficient luminosity could be obtained.
  One can also imagine some unknown
high-scale theory, which induces effective, low-energy CP-violating
interactions. The supersymmetric
extension of the Standard Model (SSM) can also produce CP violation
at the one-loop
level both in $\ttbar$ production and in the
decay~\footnote{CP violation within the SSM at future $pp$
colliders has recently been considered in Ref.~\cite{schmidt}.}.
There exists in the literature a number of papers investigating
CP violation in the
$\ttbar$ production {\it or} in the top
decay~\cite{toplit,exception}, however usually only production {\it or}
decay is considered~\footnote{Ref.~\cite{exception} is the only
exception.}. It is clear that in general CP violation may appear
in both the
production {\it and} the decay; therefore interference of
the two is crucial, as it can even lead to cancellation of both effects.
Therefore the aim of this paper is to calculate CP-violating
observables consistently, taking both sources of CP violation
into account. The problem has been solved by defining
two kinds of angular asymmetries, one sensitive to CP violation
in the production exclusively and the other in the subsequent decays.
Predictions for those asymmetries are then calculated in the
supersymmetric extension of the Standard Model, where
appropriate one-loop diagrams are induced by the gluino
exchange.

\section{The Strategy}

Recently, Peskin has proposed for $pp$ colliders
the following CP-violating asymmetry
in the production rates of $\ttbar$ pairs with CP-conjugate helicities:
\begin{equation}
\apes \equiv \frac{\#(--)-\#(++)}{({\hbox {all}}\;\ttbar)},
\label{peskin_as}
\end{equation}
where first and second entries in $(--)/(++)$
are referring to $t$
and $\bar{t}$ helicities, respectively. Since $(--)$ and $(++)$ go
into each other under CP, non-zero asymmetry
would be a sign of CP violation. An analogous quantity may be defined
for $\ttbar$ production in $\epem$ colliders as well. Here we address
the question how to define appropriate asymmetries in the $\epem$
collisions at the level of
$\ttbar$ decay products, which would be directly proportional to
Peskin's $\apes$, in other words, how to avoid a
contamination of the measurement of  $\apes$ by CP violation
in $\ttbar$ decays. Let us assume that the main decay mode of
the top quark is $\tdec$. At the tree level in the high-energy
limit the $b$ quark always has helicity $\bhel=-$,
therefore helicity conservation
tells us that $W^+$ with helicity $\wphel=0$ coming from
the top of $\thel=+$ would like to go forward in the direction of
flying top, whereas $W^+$ with $\wphel=-$ will go mainly in the opposite
direction. $W^+$ emerging from the top of $\thel=-$ with the same
helicities as above would obviously go in the opposite direction.
An analogous picture holds for the $\bar{t}$ decays. It should be noticed
that the allowed helicities of $W^+$,
at the tree level and for massless $b$ quarks,
are $\wphel=0,-$,
and that helicity conservation never
permits to produce $\wphel=+/-$ together with $\bhel=-/+$,
respectively. Therefore, it seems natural to consider the following
asymmetries:
\begin{eqnarray}
{\cal A}_{\wphel \wmhel} & \equiv &
\frac{N_{\wphel \wmhel}}{D_{\wphel \wmhel}}\\
N_{\wphel \wmhel} & \equiv & \left(
\int^{\frac{\pi}{2}}_0 d\theta \int^{\pi}_{\frac{\pi}{2}} d\bar{\theta}-
\int^{\pi}_{\frac{\pi}{2}} d\theta \int^{\frac{\pi}{2}}_0 d\bar{\theta}
\right)
\; \frac{d^2 \sigma_{\wphel \wmhel}}{d \theta\, d
\bar{\theta}}
\nonumber \\
D_{\wphel \wmhel} & \equiv &
\int^{\pi}_0 d \theta \int^{\pi}_0 d \bar{\theta}
\; \frac{d^2 \sigma_{\wphel \wmhel}}{d \theta\, d \bar{\theta}},
\nonumber
\label{asdef}
\end{eqnarray}
where $\theta,\,\bar{\theta}$ denote polar angles of $W^+,W^-$
measured in the $t,\bar{t}$ rest frames with respect to $t,\bar{t}$
directions seen from the laboratory frame,
respectively. $d^2\sigma_{\wphel \wmhel}/(d\theta\,d\bar{\theta})$
stands for the
cross-section obtained by integrating over
the full $\ttbar$ production phase space and over $W^+,W^-$ azimuthal
angles.

Hereafter we will assume that all possible CP-violating effects
for the production process $\ttprod$ can be represented by the photon
and $Z$-boson exchange in the $s$ channel.
An elegant method to calculate the asymmetries is to adopt
known~\cite{kane} helicity amplitudes for the
$\ttprod$, expressed through the most general
four form factor~\footnote{Two other
possible form factors do not contribute in the limit of zero
electron mass.}
parametrization of the vertex $V_\mu \ttbar$ ($V=\gamma$ or $Z$):
\begin{equation}
\Gamma^\mu_{Vt\bar{t}}=\frac{ig}{2}\bar{u}(p_t)
\left[\gamma^\mu(A_V-B_V\gamma_5)
+\frac{p_t-p_{\bar{t}}}{2\,m_t}(C_V-D_V\gamma_5)\right]v(p_{\bar{t}}),
\label{vttbar}
\end{equation}
where $g$ is the $SU(2)$ gauge coupling constant.
A   similar parametrization of $\tdec$ and $\tbardec$
will be adopted:
\begin{eqnarray}
\Gamma^\mu & = & \frac{-igV^{KM}_{tb}}{\surd 2}\bar{u}(p_b)\left[
\gamma^\mu (f^L_1 P_L+f^R_1 P_R)-\frac{i\sigma^{\mu \nu } k_\nu }{m_W}
(f^L_2 P_L+f^R_2 P_R)\right] u(p_t),\\
\bar{\Gamma}^\mu & = & \frac{-ig{V^{KM}_{tb}}^\ast}
{\surd 2}\bar{v}(p_{\bar{t}})\left[
\gamma^\mu (\bar{f}^L_1 P_L+\bar{f}^R_1 P_R)-
\frac{i\sigma^{\mu \nu } k_\nu }{m_W}
(\bar{f}^L_2 P_L+\bar{f}^R_2 P_R)\right] v(p_{\bar{b}}),
\label{decays}
\end{eqnarray}
where $P_{R/L}$ are projection operators, $k$ is the
$W$ momentum and $V^{KM}$ is the Kobayashi--Maskawa matrix.
Again, because $W$ is on shell,
two additional form factors do not contribute.
It is easy to show that\footnote
{We thank W.Bernreuther for pointing an error in third paper
of Ref.~\cite{toplit},
where signs of the second equation were reversed.}:
\begin{equation}
f^{L,R}_1=\pm \bar{f}^{L,R}_1,\;\;\;\;\;\;f^{L,R}_2=\pm \bar{f}^{R,L}_2,
\label{cptrans}
\end{equation}
where the upper (lower) signs are those for contributions induced by
CP-conserving \\ (-violating) interactions.

Amplitudes for $W^+W^-$ helicities $(0,0)$ and $(-+)$ could be written
in terms of amplitudes for $\ttprod$ production
$(h_{e^-},h_{e^+},h_t,h_{\bar{t}})$
and for $t$ $(h_t,\lambda_{W^+},h_b)$ and $\bar{t}$
$\overline{(h_{\bar{t}},\lambda_{W^-},h_{\bar{b}})}$
decays, as follows:
\begin{eqnarray}
M_{\wphel \wmhel} & = &
\sum_{\ephel \emhel} M^{\emhel \ephel}_{\wphel \wmhel}\\
M^{-+}_{00} & = &
(-+--)(-0-)\overline{(-0+)}+(-+-+)(-0-)\overline{(+0+)}+\nonumber \\
& & (-++-)(+0-)\overline{(-0+)}+(-+++)(+0-)\overline{(+0+)}\nonumber \\
M^{-+}_{-+} & = &
(-+--)(---)\overline{(-++)}+(-+-+)(---)\overline{(+++)}+\nonumber \\
& & (-++-)(+--)\overline{(-++)}+(-+++)(+--)\overline{(+++)}.\nonumber
\label{amplitudes}
\end{eqnarray}
Since we are interested in the leading contribution to the asymmetries
(which is an interference between one-loop and tree-level diagrams)
all other amplitudes, which in general should be included above,
have been neglected.
In order to obtain the amplitudes $M^{+-}_{\wphel \wmhel}$
one has to exchange $\ephel$ and $\emhel$ on the right-hand side of the
last two formulas.
One should notice that
the direct multiplication of helicity amplitudes without
$t$ and $\bar t$ propagators is allowed only under the so-called
narrow width approximation, which effectively means that
a $\ttbar$ pair is produced on shell and then decays.
Since $\Gamma_t\simeq 175\: \mev (\frac{m_t}{m_W})^3 \ll m_t$,
therefore the approximation is justified.

Now we are in a position to calculate the asymmetries. Adopting
helicity amplitudes from Ref.~\cite{kane}, after straightforward
algebra one gets:
\begin{eqnarray}
N_{00} & = & -\frac{8}{3} \pi^2 \left(\frac{m_t}{m_W}\right)^4
S \beta (t Q_e)^2
\sum_{i=L,R}(A^{(0)}_\gamma + r_i A^{(0)}_Z)\re\{D^{(1)}_\gamma+
r_iD^{(1)}_Z\} \\
D_{00} & = & \frac{8}{3} \pi^2 \left(\frac{m_t}{m_W}\right)^4 S (t Q_e)^2
\sum_{i=L,R}[(3-\beta^2)(A^{(0)}_\gamma + r_i A^{(0)}_Z)^2 +
2(r_i \beta B^{(0)}_Z)^2],
\label{numden00}
\end{eqnarray}
where
$$
\begin{array}{cllcll}
\beta & = & [1-\frac{4 m^2_t}{S}]^{1/2}, & t & = & 1-\frac{m^2_Z}{S},
\\
r_i & = & \frac{e_i}{t Q_e}, & & & \\
Q_e & = & -s_\theta, & Q_t & = & \frac{2}{3} s_\theta, \\
e_L & = & \frac{1}{c_\theta}(-\frac{1}{2}+s^2_\theta), &
e_R & = & \frac{1}{c_\theta} s^2_\theta,  \\
A^{(0)}_\gamma & = & 2 Q_t, & A^{(0)}_Z & = & \frac{1}{2c_\theta}
(1-\frac{8}{3} s^2_\theta), \\
B^{(0)}_Z & = & \frac{1}{2 c_\theta}. & & &
\end{array}
$$
$S$ is the total c.m. energy and $s_\theta$/ $c_\theta$ denote the sine
and cosine of the Weinberg angle, respectively;
$A^{(0)/(1)}_{\gamma / Z}$ and
$B^{(0)/(1)}_{\gamma / Z}$ stand for form factors defined in
Eq.(\ref{vttbar}), calculated at tree or one-loop level for $\gamma$ or
$Z$
vertices.
Part of common factors, corresponding to the helicity matrix element
factorization adopted in Ref.~\cite{kane}
has been dropped in $N_{00}$ and $D_{00}$.

We can see that ${\cal A}_{00}$ satisfies our expectations: {\it
it is sensitive exclusively to CP violation during production},
the decays enter at the tree level only
(and therefore they do not violate CP).
One can easy check that, assuming tree-level couplings
in the production, because of a fortunate choice for angular
integration, that all CP-violating effects from decays cancel in the
asymmetry.

It turns out that for $(\wphel \wmhel)=(-+)$ we obtain:
\begin{equation}
N_{-+}=-4\left(\frac{m_W}{m_t}\right)^4 N_{00},\; \; \;
D_{-+}= 4\left(\frac{m_W}{m_t}\right)^4 D_{00}.
\label{numdenmp}
\end{equation}
Therefore the asymmetries read:
\begin{eqnarray}
{\cal A}_{00} & = & - \beta
\frac{\sum_{i=L,R}(A^{(0)}_\gamma + r_i A^{(0)}_Z)\re\{D^{(1)}_\gamma+
r_i D^{(1)}_Z\}}
{\sum_{i=L,R}[(3-\beta^2)(A^{(0)}_\gamma + r_i A^{(0)}_Z)^2 +
2(r_i \beta B^{(0)}_Z)^2]}, \\
{\cal A}_{-+} & = & - {\cal A}_{00}.
\label{asymm}
\end{eqnarray}
Except for $(\wphel \wmhel)=(00),(-+)$, there are of course two more
configurations $(0+)$ and $(-0)$, which can also appear in the final
state even at the tree level.
However, in that case it is more appropriate to consider the following
asymmetry:
\begin{equation}
{\cal A}_{dec} \equiv \frac{N_{0+}+N_{-0}}{D_{0+}+D_{-0}}.
\label{asdec}
\end{equation}
Using a method analogous to that above
and dropping the same common factors
in numerator and denominator, one obtains:
\begin{eqnarray}
N_{0+}+N_{-0} & = & -\frac{64}{3} \pi^2 \frac{m_t(m^2_t-m^2_W)}{m_W^3}
S \beta (t Q_e)^2\re\{\bar{f}^L_2-f^R_2\} B^{(0)}_Z
\sum_{i=L,R} r_i(A^{(0)}_\gamma + r_i A^{(0)}_Z) \nonumber \\
D_{0+}+D_{-0} & = & 4 \left(\frac{m_W}{m_t}\right)^2 D_{00}.
\label{nummix}
\end{eqnarray}
{}From Eq.(\ref{cptrans}) we see that all CP-conserving
contributions cancel in the difference
$\re\{\bar{f}^L_2-f^R_2\}$  and we obtain the purely CP-violating
effect $-2\,\re\{{f^R_2}_{CP}\}$.
Therefore ${\cal A}_{dec}$ reads:
\begin{equation}
{\cal A}_{dec} = 4 \beta \left(\frac{m_t}{m_W}-\frac{m_W}{m_t}\right)
\re\{{f^R_2}_{CP}\} B^{(0)}_Z
\frac{\sum_{i=L,R}r_i(A^{(0)}_\gamma+r_iA^{(0)}_Z)}
{\sum_{i=L,R}[(3-\beta^2)(A^{(0)}_\gamma + r_i A^{(0)}_Z)^2 +
2(r_i \beta B^{(0)}_Z)^2]}.
\label{asdecres}
\end{equation}
One should notice, however, that it is not the real part of the $f^R_2$
form factor that is the sign of CP violation, but the real part
of its CP-violating component.
Now, it is obvious, that if one just na\"{\i}vely
considers the tree-level
SM-type top-quark decays and include all possible $W$
helicities,
then contributions from $(\wphel \wmhel)=(0+),(-0)$ are the ones that
contaminate CP violation in the production.
Defining ${\cal A}_{dec}$ we have not only avoided the contamination
of the measurement, but {\it also} we have found an effective method
to measure CP violation purely in the decay.
Therefore,
in order to isolate CP violation in the production and maximalize
the number of events one should
measure the following quantity:
\begin{equation}
{\cal A}_{tot} \equiv \frac{N_{00}+N_{-+}}{D_{00}+D_{-+}}
= - \frac{4m^4_W-m^4_t}{4m^4_W+m^4_t} {\cal A}_{00}.
\label{astot}
\end{equation}
However, one should also keep in mind that increasing the
top-quark mass $m_t$ the rate $\Gamma_T$ for production of
the transversal
($\wphel=-$) $W$ component is reduced with respect to
the longitudinal
($\wphel=0$) one $\Gamma_L$. According to Ref.~\cite{russel}
we have:
\begin{equation}
\frac{\Gamma_L}{\Gamma_T}=\frac{1}{2} \left(\frac{m_t}{m_W}\right)^2.
\label{gammlt}
\end{equation}
Therefore, for a very heavy top quark ($m_t \gg m_W$),
${\cal A}_{tot}$ reduces to the initial ${\cal A}_{00}$,
as independently seen from the above Eq.(\ref{astot}).

Decay modes with  $(\wphel \wmhel)=(0+),(-0)$ also provide another
interesting asymmetry obtained from that defined in
Eq.(\ref{asdef})
by a modification of the $\bar{\theta}$ integration region:
\begin{eqnarray}
\tilde{\cal A}_{\wphel \wmhel} & \equiv &
\frac{\tilde{N}_{\wphel \wmhel}}{D_{\wphel \wmhel}}\\
\tilde{N}_{\wphel \wmhel} & \equiv & \left(
\int^{\frac{\pi}{2}}_0 d\theta \int^{\frac{\pi}{2}}_0 d\bar{\theta}-
\int^{\pi}_{\frac{\pi}{2}} d\theta \int^\pi_{\frac{\pi}{2}}d\bar{\theta}
\right)
\; \frac{d^2 \sigma_{\wphel \wmhel}}{d \theta\, d \bar{\theta}},
\nonumber
\label{newas}
\end{eqnarray}
for $(\wphel \wmhel)=(0+),(-0)$. The result is:
\begin{equation}
\tilde{\cal A}_{-0}=-{\cal A}_{00}\, , \; \; \;
\tilde{\cal A}_{0+}=+{\cal A}_{00},
\label{newasres}
\end{equation}

\section{The Asymmetries in a Supersymmetric Standard Model}

Both ${\cal A}_{\wphel \wmhel}$ and ${\cal A}_{dec}$ are induced by terms
that cannot be present at tree level in any renormalizable theory.
Therefore they may appear either as an artifact of some unknown
high-scale theory in an effective low-energy Lagrangian or they can
be generated at the loop level in the SM or its extensions.
In the SM both $D$ and ${f^R_2}_{CP}$
vanish, even at the one-loop
level; we therefore illustrate the above general consideration
by the one-loop-generated $D$ in the SSM.
For our purpose the most relevant
new source of CP violation, which appears in the SSM, would be the phase
of gluino ($\lambda^a$), top quark and top squark ($\tilde t$)
coupling:
\begin{equation}
{\cal L} = i\surd 2 g_s
[e^{i\phi} {\tilde t}^\ast_L T^a (\bar {\lambda}^a t_L)+
e^{-i\phi} {\tilde t}^\ast_R T^a (\bar {\lambda}^a t_R)] + \hc,
\label{glulag}
\end{equation}
where $g_s$ is the QCD coupling constant.
The stop quarks of different handedness are related to
the stop-quark mass eigenstates ${\tilde t}_1$, ${\tilde t}_2$
through the following rotation:
\begin{eqnarray}
\sstop_L & = & \sstop_1 \cos \alpha + \sstop_2 \sin \alpha \nonumber \\
\sstop_R & = &-\sstop_1 \sin \alpha + \sstop_2 \cos \alpha
\label{mix}
\end{eqnarray}
As we will see, CP-violating
effects vanish for degenerated $\sstop_{1/2}$
masses and are proportional to the product $\sin 2\alpha \sin 2\phi$.
However, the same quantity appears in the formula for the neutron's
electric dipole moment (EDM); we therefore have to take
into account limits originating  from the EDM measurement.
As estimated in Ref.~\cite{schmidt} for $\sin 2\alpha \sin 2\phi=1$
the maximal allowed splitting between stop-quark masses is roughly
$400 \: \gev$, we will satisfy this bound in our
calculations.\footnote{
It should be stressed here that we are not restricting ourselves
to the minimal supergravity-induced models. Therefore the bounds
on CP-violating parameters from  neutron's EDM measurements
are rather weak. However, as showed in Ref.~\cite{kizukuri}, even
the supersymmetric Standard Model based on supergravity
allows for maximal CP-violating phases for sufficiently
heavy squarks, therefore it is legitimate to use $\phi=\pi/4$.}
The result for $\re \{ D_{\gamma/Z} \} $ is the following:
\begin{eqnarray}
\re\{D_\gamma\} & = & - \delta_{CP} \frac{8s_W}{3\pi \beta^2}
[\im(1,1)-\im(2,2)] \nonumber \\
\re \{ D_Z \} & = & \delta_{CP}\frac{2}{c_W \pi \beta^2} \\
& & \left[\left(\cos^2 \alpha - \frac{4}{3} s^2_W \right) \im(1,1)  -
                          \cos 2 \alpha \im(1,2)  -
   \left(\sin^2 \alpha-\frac{4}{3}s^2_W\right)\im(2,2)\right],\nonumber
\label{realparts}
\end{eqnarray}
where
\begin{equation}
\delta_{CP}=\alpha_{QCD}\frac{m_\lambda m_t}{S} \sin 2\alpha \sin 2\phi
\label{delta}
\end{equation}
and $\im(ij)$ (given in the Appendix)
corresponds to the exchange of $\sstop_i,\sstop_j$
in the loop (see Fig.~\ref{diag}).

First, we assume that $\delta_{CP}$ can reach its maximum for
$\alpha=\phi=\pi/4$, and then we adjust squark masses to obtain
$\re\{D\}$ as big as possible. Notice that for such angles there
is no mixed $(1,2)$ contribution in the loop.
Since we need an imaginary part for the diagram (see Appendix),
 in order to maximalize
the effect we choose
one stop quark ($\sstop_1$) as light as experimentally allowed:
$\sstop_1$ can be as light as $m_Z/2$~\cite{stop}, so
we fix its mass
$m_1$ at $50\:\gev$ and vary $m_2$, the mass of the remaining stop.
It is clear that CP violation vanishes if $m_1=m_2$, therefore
the biggest contribution can be obtained if $m_2$ is shifted
above the threshold for $\sstop_2$ production. Our results
correspond to $m_2 = 250\: \gev$, which is of course allowed
by bounds from Ref.~\cite{schmidt}.


Two diagrams ($V=\gamma,Z$)
contributing to $D_V$ are shown in Fig.~\ref{diag}.

It is easy to notice that there are no contributions to $D$ from
self-energy diagrams.
As seen from the formula for $\re\{D_{\gamma/Z}\}$ only an absorptive
part of the vertex is necessary.


The results for ${\cal A}_{00}$ at future high-energy $\epem$
linear collider of $\surd S =500\:\gev$ are presented
in Fig.~\ref{asymm_00}. For the gluino mass we use $m_\lambda=200.\:\gev$.
The maximal asymmetry is about $0.6\:\%$.


As discussed above,      in order to maximalize the number of events
one should measure ${\cal A}_{tot}$, which we plot
in Fig.~\ref{asymm_to}. The            biggest asymmetry
is about $0.5\:\%$. For the most optimistic proposal of $500\: \gev$
linear collider~\cite{topgroup}
the total integrated luminosity is $ 8.5 \times 10^4 \: \lumun $,
therefore the SM cross-section for $\ttbar$ production ($0.66\:\pbarn$
at $m_t=150\: \gev$)~\cite{topgroup} predicts about $N\simeq 5000$ events
per year.
This means that the smallest possible measurable asymmetry is about
$1/\surd N=0.4\:\%$. One should also have in mind that the above
estimates do not include any cuts and, of course, some number of events
must be lost because of non-perfect efficiency.
The results that we have obtained here presumed the narrow width
approximation, where all possible interference effects between
production and decay are neglected. In order to justify this
we must in addition
assume that the final $(Wb)$ mass resolution is sufficiently
good to be sure that $W$ and $b$ are coming from on-shell top quarks.
There is still a problem of identification of the $W$ helicities.
However, it is useful to notice here that, because of helicity
conservation, $e^+$ coming from $W^+$ with $\wphel=0$ would like to
go either along the $W$ direction or opposite to it with the same
probability.
In the case of $\wphel=-$, $e^+$ would strongly
prefer to fly against the original $W$ direction. That observation
should be helpful in the process of $W$ helicity identification;
however, any qualitative analysis would require a dedicated Monte
Carlo, which is obviously beyond the scope of this paper.

We must
conclude that the maximal SSM asymmetries will be just at the edge
of possible observability
at planned $\surd S =500\:\gev$ colliders.

The general consideration presented above assume that the main
$t$ decay mode is $\tdec$, which may not be true in extensions of
the SM. Also our example within the SSM needs the extra assumption
that, in spite of very light $\sstop_1$, $\tdec$ still dominates.
This means that zino ($\tilde{Z}$) and photino ($\tilde{\gamma}$)
must be sufficiently heavy to forbid otherwise \nopagebreak
competitive
decays: $t \rightarrow \sstop_1 (\tilde{Z} / \tilde{\gamma})$.

It is worthwhile
 to emphasize here that the asymmetries in the SSM are induced by
the strong interaction (gluino exchange) and it is therefore hard to
imagine that any bigger asymmetry would appear at the one-loop level.

In Fig.~\ref{asymm_s} we show the ${\cal A}_{00}$ dependence
on $\surd S$ for different top-quark masses. It is amusing to notice
that the biggest asymmetry is always obtained for $\surd S$ close to
$500 \: \gev$.

One more comment is needed here, namely the asymmetries considered
in this paper are
really CP violation observables; one can check, for instance by direct
calculation, that gluonic and photonic vertex corrections
do not generate any contributions to the asymmetries,
even though
they develop non-zero absorptive parts.

The gluino exchange contributes also to the $\tdec$ and therefore
one may calculate ${f^R_2}_{CP}$ at the one-loop level.
However, because in this case also the bottom sector
enters, the result would involve too many unconstrained parameters.
Therefore that calculation will not be presented here.

\section{Summary}

CP violation in the production {\it and} decay of $\ttbar$
at future high energy linear colliders has been discussed
in a general model-independent way. Since experimentally one can never
separate the top-quark production and its decay, the two mechanisms
must be considered simultaneously.
In order to avoid mixing of
CP violation in the production and the decay we have defined
the amplitudes sensitive to one or the other mechanism.
It turns out that when the asymmetries are
induced by gluino exchange the maximal asymmetry will be just
at the level of observability, assuming the most optimistic
scenario.
Nevertheless, they should be searched
experimentally, as their possible measurement could also be a signal
of some high scale, unknown CP-violating theory.


\appendix
\section*{Appendix}
The function $\im(i,j)$ is defined as follows:
\begin{eqnarray}
\im(i,j) & = & \im\{ B_0(p_{\bar t};m_\lambda^2,m_j^2)
+B_0(-p_t;m_\lambda^2,m_i^2)-2B_0(p_t+p_{\bar t};m_i^2,m_j^2) +
\nonumber \\
& & -[m_i^2+m_j^2-S-2(m_\lambda^2-m_t^2)] C_0(-p_t,p_{\bar t};
m^2_\lambda,m_i^2,m_j^2) \}. \nonumber
\label{imfun}
\end{eqnarray}
$B_0$ and $C_0$ are standard 2- and 3-point scalar one-loop integrals:
\begin{eqnarray}
& \frac{i}{16\pi^2}B_0(k;m_1^2,m_2^2) & \equiv \mu^{4-n} \int
\frac{d^n r}{(2\pi)^n} \frac{1}{[r^2-m_1^2+i\varepsilon]
[(r+k)^2-m_2^2+i\varepsilon]} \nonumber \\
& \frac{-i}{16\pi^2}C_0(k,p;m_1^2,m_2^2,m_3^2) & \equiv \nonumber \\
\lefteqn{
\mu^{4-n} \int
\frac{d^n r}{(2\pi)^n} \frac{1}{[r^2-m_1^2+i\varepsilon]
[(r+k)^2-m_2^2+i\varepsilon]
[(r+p)^2-m_3^2+i\varepsilon]}},\nonumber
\label{integ}
\end{eqnarray}
where $\mu$ is the regularization scale.
Imaginary parts of those integrals are, of course, finite and
are given below
for the relevant configurations:
\begin{eqnarray}
\im\{B_0(k;m^2,m^2)\} & = & \frac{\pi}{k^2} w(k^2,m^2,m^2)
\Theta(k^2-4m^2) \nonumber \\
\im\{C_0(-p_t,p_{\bar t};m_\lambda^2,m_t^2,m_t^2)\} & = &
\frac{\pi}{S \beta} \log\left[
\frac
{m^2_1+m^2_2+2(m^2_t-m^2_\lambda)-S-\beta w(S,m_1^2,m_2^2)}
{m^2_1+m^2_2+2(m^2_t-m^2_\lambda)-S+\beta w(S,m_1^2,m_2^2)}
\right] \nonumber \\
& & \Theta[S-(m_1+m_2)^2], \nonumber
\label{imparts}
\end{eqnarray}
where $w(a,b,c)$ is the standard kinematic function:
\begin{eqnarray}
w(a,b,c) & = & [(a+b-c)^2-4ab]^{1/2}.
\nonumber
\end{eqnarray}


\end{document}